\documentclass[10pt,superscriptaddress,twocolumn,amsmath,amssymb,aps,pra]{revtex4}
\usepackage{epstopdf}
\usepackage{mathrsfs}
\usepackage{graphicx}% Include figure files
\usepackage{dcolumn}% Align table columns on decimal point
\usepackage{bm}% bold math

\newcommand{\be}{\begin{equation}}
\newcommand{\ee}{\end{equation}}
\newcommand{\bea}{\begin{eqnarray}}
\newcommand{\eea}{\end{eqnarray}}

\begin{document}
\title{Meroniums in Spin-Orbit Coupled Bose Gases}
\author{Bo Chen}
\affiliation{Institute of Theoretical Physics, School of Physics and Optoelectronic Engineering, Beijing University of Technology, Beijing, 100124, China}
\author{Xiaojing Jin}
\affiliation{Institute of Theoretical Physics, School of Physics and Optoelectronic Engineering, Beijing University of Technology, Beijing, 100124, China}
\author{Jiatao Tan}
\affiliation{Institute of Theoretical Physics, School of Physics and Optoelectronic Engineering, Beijing University of Technology, Beijing, 100124, China}
\author{Boyang Liu}\email{boyangleo@gmail.com}
\affiliation{Institute of Theoretical Physics, School of Physics and Optoelectronic Engineering, Beijing University of Technology, Beijing, 100124, China}

\date{\today}
\begin{abstract}
In this work we demonstrate the existence of new types of topological states in a two-component Bose gas with Rashba spin-orbit couplings. We construct a wave function by mapping the degenerate ground states to the circular boundary of the system, and further consider its superposition with the time-reversed counterpart. By employing the imaginary-time evolution method, we identify four types of topological phases. The first one is the well-known half vortex(HV). The second one is called radial wave half vortex(RWHV), which is a combination of HV with a radial wave factor $e^{ikr}$. The spin configurations show that both HV and RWHV are merons. The third and fourth types are of particular interest. The third one is a combination of HV and anti-HV. The spin densities of this phase demonstrates peak structures and the peaks of spin up and spin down densities are mismatched, and hence we call it the double peak (DP) phase. The fourth type is a combination of RWHV and anti-RWHV, due to the superposition of the radial wave factor $e^{ikr}$ and $e^{-ikr}$ the spin densities demonstrate a spiral pattern, then it's called spin spiral (SS) phase. They are both combinations of meron and antimeron. Here, we name them as meroniums. The topological charges of the four phases are calculated and the spin distributions are demonstrated.
\end{abstract}
 \maketitle
\section{Introduction}
Driven by both fundamental interest and technological applications, the search for novel topological states has been a frontier of condensed matter physics. Well known examples are two dimensional spin textures skyrmions and merons. Skyrmion is a topological field configuration that was first proposed to describe nucleon by Tony Skyrme in 1961  \cite{Skyrme1961}. Since then, the analogous topological spin texture have been studied in many contexts \cite{Ho1998,Ohmi1998,Khawaja2001,Leslie2009,Choi2012,Sondhi1993,Barrett1995,RObler2006,Muhlbauer2009,Fert2017,Bogdanov2020,Tokura2021,Fernandes2022,Tsesses2018,Du2019,Mechelen2019}. Meron is topologically equivalent to one-half of a skyrmion since the topological charge of it is $\pm \frac{1}{2}$.  It has been theoretically investigated \cite{Lin2015,Kharkov2017,Gobel2019,Zhang2020,Guo2021} and experimentally observed in various materials, for instance, confined magnetic disk\cite{Shinjo2000,Wintz2013}, continuous magnetic film\cite{Waeyenberge2006,Yu2018,Gao2019}, and van der Waals magnetic crystal\cite{Lu2020,Augustin2021}.  Due to the topological features and small sizes these spin textures can have potential applications in information storage and processing.

The combinations of skyrmions or merons are also intriguing objects and have attract enormous interest. A skyrmion and anti-skyrmion can form a pair, the so-called skyrmionium \cite{Zhang2016, Kolesnikov2018, Hagemeister2018, Zhang2018, Bo2020, Obadero2020, Seng2021, Tang2021, Yang2023}, with a zero topological charge. This state is of particular interest due to its lacking of skyrmion Hall effect\cite{Kolesnikov2018, Tang2021}. An isolated meron in a non-confined system is usually in absence.  The spins of the meron at the core region point out of the plane, while the spins align in the plane at the periphery and swirl along the periphery like a vortex. Such a configuration is not energetically favorable. Hence, they are likely to form pairs or groups, so that their swirling spin textures can be mutually cancelled away from the cores and finally results in localized spin configurations with finite energy. A bimeron is a combination of two merons \cite{Komineas2007, Zhang2015, Heo2016, Kharkov2017, Shen2020}, resulting in a net integer topological charge (usually $\pm 1$). Recently, a new state named as bimeronium is experimentally observed \cite{Zhang2021}, which is a combination of two bimerons with opposite topological charges, resulting in a zero net topological charge. Cold atom systems provide an indispensable platform for quantum simulation, owing to their remarkable tunability and rich tools of experimental probes. Notably, various types of topological structures have been created in cold atom experiments, including skyrmions\cite{Choi2012}, monopoles\cite{Ray2014,Ray2015}, and knots\cite{Hall2016}, demonstrating the capacity of the cold atom systems to explore fundamental physics beyond traditional condensed matter settings. Therefore, it is intriguing to explore whether such a topological state with zero net topological charges exists in cold atomic systems.
%Similar to the skyrmion, the spins of the meron at the core region point out of the plane in the two-dimensional system. However, the spin structure of merons are different from the skyrmions at the periphery. For merons the spins align in the plane, while they point out of the plan for the skyrmions as shown in Fig. . In this sense, an isolated meron is not energy favorable in a continuous system. They intend to form pairs or groups, so that their swirling spin textures can be mutually cancelled away from the cores and finally results in localized spin configurations with finite energy.

In this work we investigate the topological states in a two-dimensional Rashba spin-orbit (SO) coupled Bose gas and find four different topological states exist. The HV phase has been investigated in this system \cite{Wu2011,Ramachandhran2012}. However, these investigations didn't consider an important factor of radial wave $e^{ikr}$. Our results show that such a factor can help to reduce the energy. Hence, in some parameter space, there is a so-called RWHV phase exist, which is a combination of HV and radial wave factor $e^{ikr}$. Furthermore, a HV and an anti-HV can combine to form a DP state, while a RWHV and an anti-RWHV can combine to form SS state. The spin configurations show that HV and RWHV are merons, while the DP and SS are meron-antimeron pairs, which have zero topological charges. They are named as meroniums.

Our work is organized as the following. In Sec. II we construct the wave functions of the topological states base on the analysis of the symmetry on the ground state manifold. In Sec. III we numerically solve the wave functions and study the spin density properties of the four topological phases. The phase diagram with respect to different SO coupling and interaction strengths is presented. In Sec. IV we study the spin configurations and calculate the topological charges of the topological states, and show that there exists  a new type of topological state, the meronium. Finally, Sec. V provides our conclusions.

\section{MEAN-FIELD MODEL OF THE TOPOLOGICAL STATES}
We consider a two-component Bose gas with Rashba SO coupling in a 2D harmonic trap. The Hamiltonian of the system can be written as ($\hbar=1$)
\bea
\hat H=\hat H_0+\hat H_{\rm int},
\eea
where
\bea
\hat H_{0}=\int d^2 {\bf r}\hat\Psi^\dagger({\bf r})\Big(-\frac{\nabla^{2}}{2m}+V_{\rm SOC}+V({ r})\Big)\hat\Psi(\bf r) \label{eq:H0}
\eea
is the single particle part. The spinor $\hat \Psi({\bf r})=[\hat\psi_\uparrow({\bf r}),\hat\psi_\downarrow({\bf r})]^T$ ($\hat \Psi^\dagger({\bf r})=[\hat\psi^\ast_\uparrow({\bf r}),\hat\psi^\ast_\downarrow({\bf r})]$) is the annihilation (creation) operator for the two-component Bose gas. The system is confined in a 2D harmonic trap, and the trap potential is given by $V({ r})=m\omega^2 r^2/2$. The Rashba spin-orbit coupling term is written as $V_{\rm SOC}=\frac{ik_0}{m}(\sigma_x\partial_x+\sigma_y\partial_y)$, where $k_0$ is the parameter that describes the SO coupling strength and $\sigma_i (i=x,y)$ is the Pauli matrix. The interaction part of the Hamiltonian is as the following,
\bea
\hat H_{\rm int}=\int d^2 {\bf r}\Big(g\sum_{\sigma=\uparrow,\downarrow}\hat\psi^\dagger_\sigma\hat\psi^\dagger_\sigma\hat\psi_\sigma\hat\psi_\sigma+2g^\prime\hat\psi^\dagger_\uparrow
\hat\psi^\dagger_\downarrow\hat\psi_\downarrow\hat\psi_\uparrow\Big).
\eea Here we assume equal intraspecies interaction strength $g$ for spin-up and spin-down components. The interspecies interaction strength is denoted as $g^\prime$.

It has been calculated the single particle ground state of Hamiltonian of Eq.(\ref{eq:H0}) is on a circle in the momentum space with $\sqrt{k_x^2+k^2_y}=k_0$. Furthermore, due to the SO coupling, this ground state manifold possesses a combined symmetry in both real space and spin space as the following
\bea
\left[\begin{array}{c}x\\y\end{array}\right]\rightarrow \left[\begin{array}{cc}\cos\theta & -\sin\theta\\ \sin\theta & \cos\theta\end{array}\right]\left[\begin{array}{c}x\\y\end{array}\right]~~~ {\rm and} ~~~\sigma_i\rightarrow u \sigma_i u^\dagger, \label{eq:trans}
\eea where the unitary matrix is written as
\bea u=\left[\begin{array}{cc}\cos(\theta/2)+i\sin(\theta/2) & 0\\ 0 & \cos(\theta/2)-i\sin(\theta/2)\end{array}\right].
\eea
If we choose the momentum as ${\bf k}$, where $|{\bf k}|=k_0$ and $\alpha_{\bf k}$ is the azimuth angle of ${\bf k}$ on the $xy$-plane, the ground state can be expressed in a form of plane wave $\frac{1}{\sqrt2}e^{i{\bf k}\cdot{\bf x}}\left[\begin{array}{c}e^{-i\alpha_{\bf k}/2}\\e^{i\alpha_{\bf k}/2}
\end{array}\right]$. With different interactions the system can condense into one of the ground states, the so-called plane wave phase, or a pair of ground state with opposite momenta${\bf k}$ and $-{\bf k}$, the so-called stripe phase \cite{Wang2010,Zhai2015}.

In order to construct a topological state we consider map the degenerate ground states to the circular boundary of the system. Hence, we consider a wave function of the topological state in the form of $e^{ikr}\left[\begin{array}{c}f(r)e^{-i\theta/2}\\g(r)e^{i\theta/2}\end{array}\right]$. However, there is a problem of this wave function. As the angle $\theta$ goes from $0$ to $2\pi$, that is, the system takes a whole revolution along the boundary, the spin part won't return to it original state. This is a manifest of that the group $SU(2)$ is a double cover of $SO(3)$. To fix this, we take into account total phase and assume that it swirls along the boundary with a winding number of $n+\frac{1}{2}$. Thus the wave function can be cast as
\bea
\phi_n&=&e^{ikr}e^{i(n+\frac{1}{2})\theta}\left[\begin{array}{c}f(r)e^{-i\theta/2}\\g(r)e^{i\theta/2}\end{array}\right]\cr&=& e^{ikr}\left[\begin{array}{c}f(r)\\g(r)e^{i\theta}
\end{array}\right]e^{in\theta},\label{eq:wf}\eea
This state has a well-defined total angular momentum $j_z=l_z+s_z=n+1/2$, where $n=0,1,2...$ is the angular momentum quantum number. The normalization condition is $\int d^2 {\bf r}(f(r)^2+g(r)^2)=1$. Here we need to emphasize that there is a two-dimensional radial wave factor $e^{ikr}$ in the wave function Eq.(\ref{eq:wf}), which is analogues to the ``plane wave phase" of the SO coupled system, where the BEC condenses at the state with finite momentum. So it's reasonable to have a vortex solution with a radial wave term.
In a SO coupled system, such a state doesn't carry particle currents.  According to Noether's theorem, the $U(1)$ gauge symmetry induces conserved currents $J_x=\frac{1}{2mi}\sum_{\sigma=\uparrow,\downarrow}(\hat \psi^\dagger_\sigma\partial_x\hat\psi_\sigma-\partial_x\hat \psi^\dagger_\sigma\hat\psi_\sigma)-\frac{k_0}{m}(\hat\psi^\dagger_\uparrow\hat\psi_\downarrow+\hat\psi^\dagger_\downarrow\hat\psi_\uparrow)$, $J_y=\frac{1}{2mi}\sum_{\sigma=\uparrow,\downarrow}(\hat \psi^\dagger_\sigma\partial_y\hat\psi_\sigma-\partial_y \hat \psi^\dagger_\sigma\hat\psi_\sigma)+\frac{ik_0}{m}(\hat\psi^\dagger_\uparrow\hat\psi_\downarrow-\hat\psi^\dagger_\downarrow\hat\psi_\uparrow)$. For the plane wave phase $\frac{1}{\sqrt2}e^{i{\bf k}\cdot{\bf r}}\left[\begin{array}{c}e^{-i\alpha_{\bf k}/2}\\e^{i\alpha_{\bf k}/2}\end{array}\right]$  the SO coupling term in $J_{x/y}$ exactly cancels the gradiant term, resulting in zero current. Our trial wave function is constructed through the superposition of all plane wave states in the degenerate circle and hence it doestn't carry a radial current.

\section{THE CLASSIFICATION OF THE TOPOLOGICAL STATES}
In the following discussions we will only consider the case of $n=0$. The states of $n\neq0$ are usually with higher energy. Please refer to the appendix A for the detailed energy functional of $n\neq0$ case. Since our primary goal is to demonstrate the existence of the novel topological states, restricting to the lowest-energy sector is sufficient.  Using the wave function of Eq.(\ref{eq:wf}) the mean-field energy functional of $n=0$ state can be calculated as the following
\begin{align}
&E(\phi_0)=N\int d^2 {\bf r}\Big\{\frac{1}{2m}\big[(\partial_r f(r))^2+(\partial_r g(r))^2\cr &+k^2f^2(r)+(k^2+1/r^2)g^2(r)-4kk_0f(r)g(r)\big]\cr &+\frac{1}{2}m\omega^2r^2(f^2(r)+g^2(r))+(g+g^\prime)\frac{N}{2}(f^2(r)+g^2(r))^2\cr&+(g-g^\prime)\frac{N}{2}(f^2(r)-g^2(r))^2\Big\}, \label{eq:E0}
\end{align}
where $N$ is the particle number of the bosons. It's worth to notice that there is a term of $-k k_0f(r)g(r)$, the negative sign of which indicates that a finite $k$ will probably help to lower the energy, hence, it's reasonable to have a radial wave factor $e^{ikr}$ in the wave function Eq. \ref{eq:wf}. Since the system is in a harmonic trap, we will use the trap units in this work, that is, we take $\omega$ (here we already take $\hbar=1$) as the energy scale and $a_{\perp}=\sqrt{\frac{1}{m\omega}}$ as the length scale. Then, the SO coupling strength $k_0$ and parameter $k$ are both in units of $1/a_{\perp}$, and the interaction strength $g$ and $g^\prime$ are in units of $\omega a_{\perp}^2/N$.

\begin{figure}[t]
\includegraphics[width=0.5\textwidth]{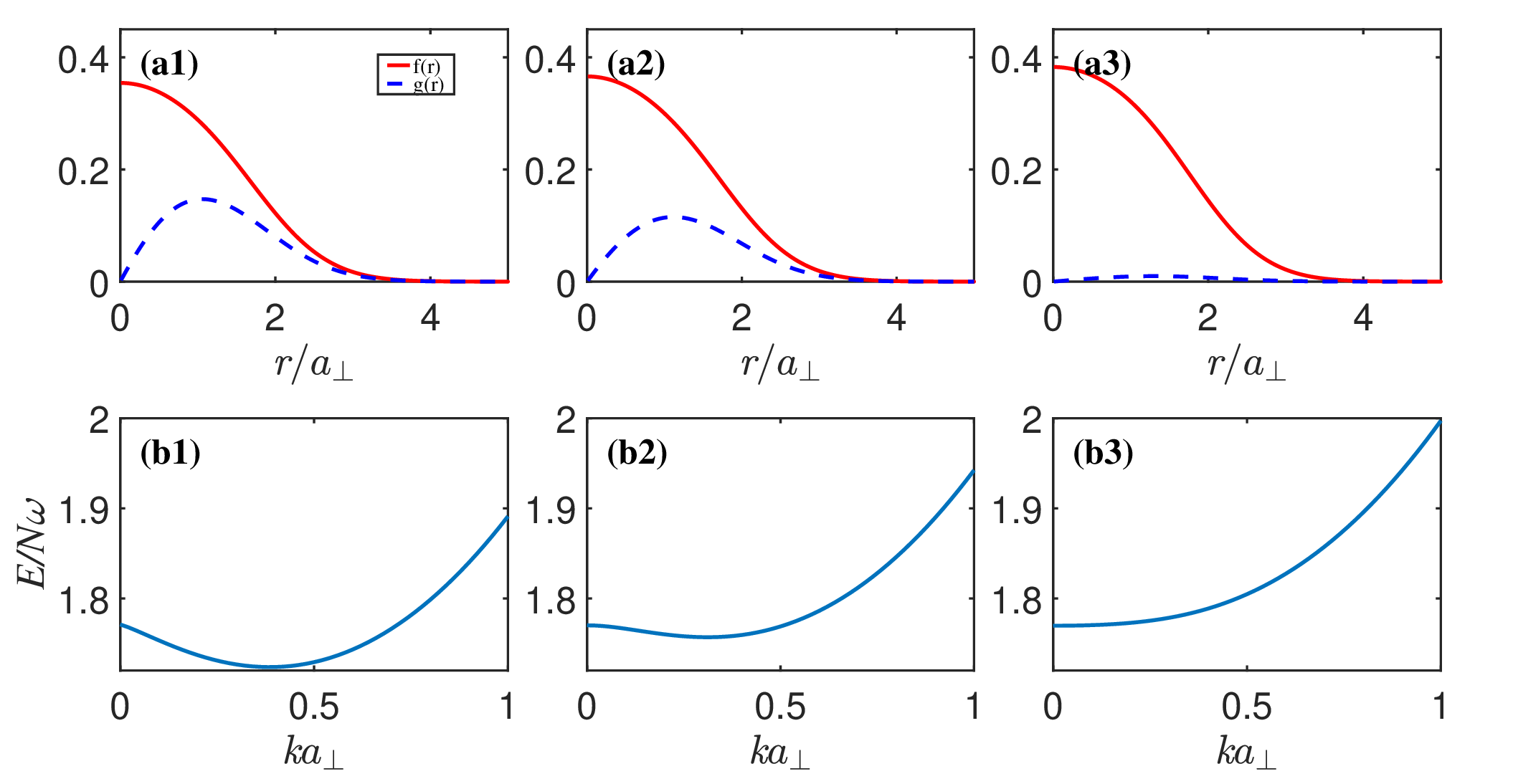}
\caption{(Color online) (a1), (a2) and (a3) demonstrate the functions of $f(r)$ and $g(r)$ with different interaction strengths. (b1), (b2) and (b3) illustrate the energy $E(\phi_0)$ as a function of $k$.  $k_0 a_{\perp}=0.5$ and $Ng/\omega a_\bot^2=7$ for all the graphs.  $Ng^\prime/\omega a_\bot^2=2.6$ for (a1) and (b1); $Ng^\prime/\omega a_\bot^2=4.4$ for (a2) and (b2); $Ng^\prime/\omega a_\bot^2=6.6$ for (a3) and (b3).}
\label{fig:fg}
\end{figure}

Minimization of the energy functional $E(\phi_0)$ yields the partial differential equations of $f(r)$ and $g(r)$ as the followings
\begin{align}
&\frac{1}{2m}\Big(-\frac{\partial^2 f(r)}{\partial r^2}-\frac{1}{r}\frac{\partial f(r)}{\partial r}+k^2f(r)\Big)-\frac{kk_0}{m}g(r)\cr &+\frac{1}{2}m\omega^2r^2f(r)+N(g+g^\prime)\big(f^2(r)+g^2(r)\big)f(r)\cr&+N(g-g^\prime)\big(f^2(r)-g^2(r)\big)f(r)-\mu f(r)=0, \cr&\frac{1}{2m}\Big(-\frac{\partial^2 g(r)}{\partial r^2}-\frac{1}{r}\frac{\partial g(r)}{\partial r}+\big(k^2+\frac{1}{r^2}\big)g(r)\Big)-\frac{kk_0}{m}f(r)\cr &+\frac{1}{2}m\omega^2r^2g(r)+N(g+g^\prime)\big(f^2(r)+g^2(r)\big)g(r)\cr&+N(g-g^\prime)\big(f^2(r)-g^2(r)\big)g(r)-\mu g(r)=0,\label{eq:EL}
\end{align}
Please refer to the appendix A for the detailed derivation of the energy functional and above Euler-Lagrange equations. For different parameters $k_0$, $g$, $g^\prime$,  the functions of $f(r)$ and $g(r)$ can be numerically solved. In Fig. \ref{fig:fg} (a1),(a2) and (a3) we plot the function $f(r)$ and $g(r)$ for different interactions. With the solution of $f(r)$ and $g(r)$ the energy $E(\phi_0)$ can be calculated and we plot $E(\phi_0)$ as a function of $k$ in Fig. \ref{fig:fg} (b1),(b2) and (b3). One observes that the energy minimum can be located at either $k=0$ or $k\neq k_0$ for different interaction strengths. We use $k_c$ to denote the location of the energy minimum.  In all the graphs of Fig. \ref{fig:fg} the intraspecies interaction $g$ is fixed while the interspecies interaction $g^\prime$ are different. In Fig. \ref{fig:fg} (b1) the interaction strength is $Ng^\prime/\omega a_\bot^2=2.6$. We find that the energy minimum is located at $k_c\approx k_0$. As the interspecies interaction increases to $Ng^\prime/\omega a_\bot^2=4.4$ the energy minimum is lift up, and the location of the minimum approaches zero as shown in Fig.\ref{fig:fg} (b2). As the interspecies interaction is further increased to $Ng^\prime/\omega a_\bot^2=6.6$ we see that energy minimum is eventually located at the point $k=0$ as show in Fig.\ref{fig:fg} (b3). Hence, we find that $k_c$ could be either zero or finite. There is a second order phase transition as the interaction strengths vary.

To numerically solve differential equations of Eq.(\ref{eq:EL}), we employ the imaginary-time evolution method with a semi-implicit Crank-Nicolson scheme and a time step $\Delta \tau=10^{-4}$, iterating until the maximum absolute difference between successive wave functions falls below the convergence tolerance $10^{-6}$. The radial coordinate is discretized on a uniform grid with spacing $\Delta r/a_\perp=0.01$ in the region $r/a_\perp\in[0,10]$ . At the inner boundary ($r=0$), a condition $\partial_r f(r)=0$ is imposed to ensure a smooth profile, while the function $g(r)$ satisfies a condition of $g(r)=0$ due to the strong centrifugal barrier $1/r^2$. At the outer boundary, the harmonic trap guarantees $f(r)\rightarrow0$ and $g(r)\rightarrow0$. To determine the critical momentum $k_c$, we treat $k$ as a variational parameter in the range $0\leq k\leq 2k_0$. For each fixed $k$, we converge to the corresponding stationary solution and evaluate the energy functional $E(k)$, with $k_c$ identified as the global minimum of $E(k)$. Finally, the phase diagrams are constructed by scanning the interaction strengths over $Ng/\omega a_\bot^2,Ng^\prime/\omega a_\bot^2\in[0,14]$ with a mesh step $N \Delta g/\omega a_\bot^2=N \Delta g^\prime/\omega a_\bot^2=0.2$.  The phase boundaries are determined according to the criteria $k_c=0$ vs. $k_c\neq0$ and the sign of the superposition energy difference $\Delta E$.

Furthermore, the Hamiltonian $\hat H_0$ is invariant under the time-reversal symmetry, hence, there is a degenerate time-reversal state, which is written as
\bea && \phi_0^{\mathcal T}=\mathcal T \phi_0= e^{-ikr}\left[\begin{array}{c}g(r)e^{-i\theta}\\-f(r)
\end{array}\right].
\eea
It's interesting to explore the superposition state of $\phi_0$ and $\phi_0^{\mathcal T}$, which can be written as
\bea
\phi_s=\alpha \phi_0+\beta\phi_0^{\mathcal T}, \label{eq:phis}
\eea where $|\alpha|^2+|\beta|^2=1$.
Due to the interaction, such a superposition state will have an energy $E(\phi_s)$ different from $E(\phi_0)$. Straight-forward calculation shows that
\bea
&&\Delta E=E(\phi_s)-E(\phi_0)=2N^2(g^\prime-g)|\alpha|^2|\beta|^2\cr&&\int d^2 {\bf r}\big\{[f^2(r)-g^2(r)]^2-2f^2(r)g^2(r)\big\}.
\eea
If $\Delta E>0$ the system condenses into the state $\phi_0$ or $\phi_0^{\mathcal T}$, that is, $\alpha=1,\beta=0$ or $\alpha=0,\beta=1$, respectively. Otherwise, the system condenses into a superposition state of $\alpha=\beta=1/\sqrt 2$.

\begin{figure}[h]
\includegraphics[width=0.5\textwidth]{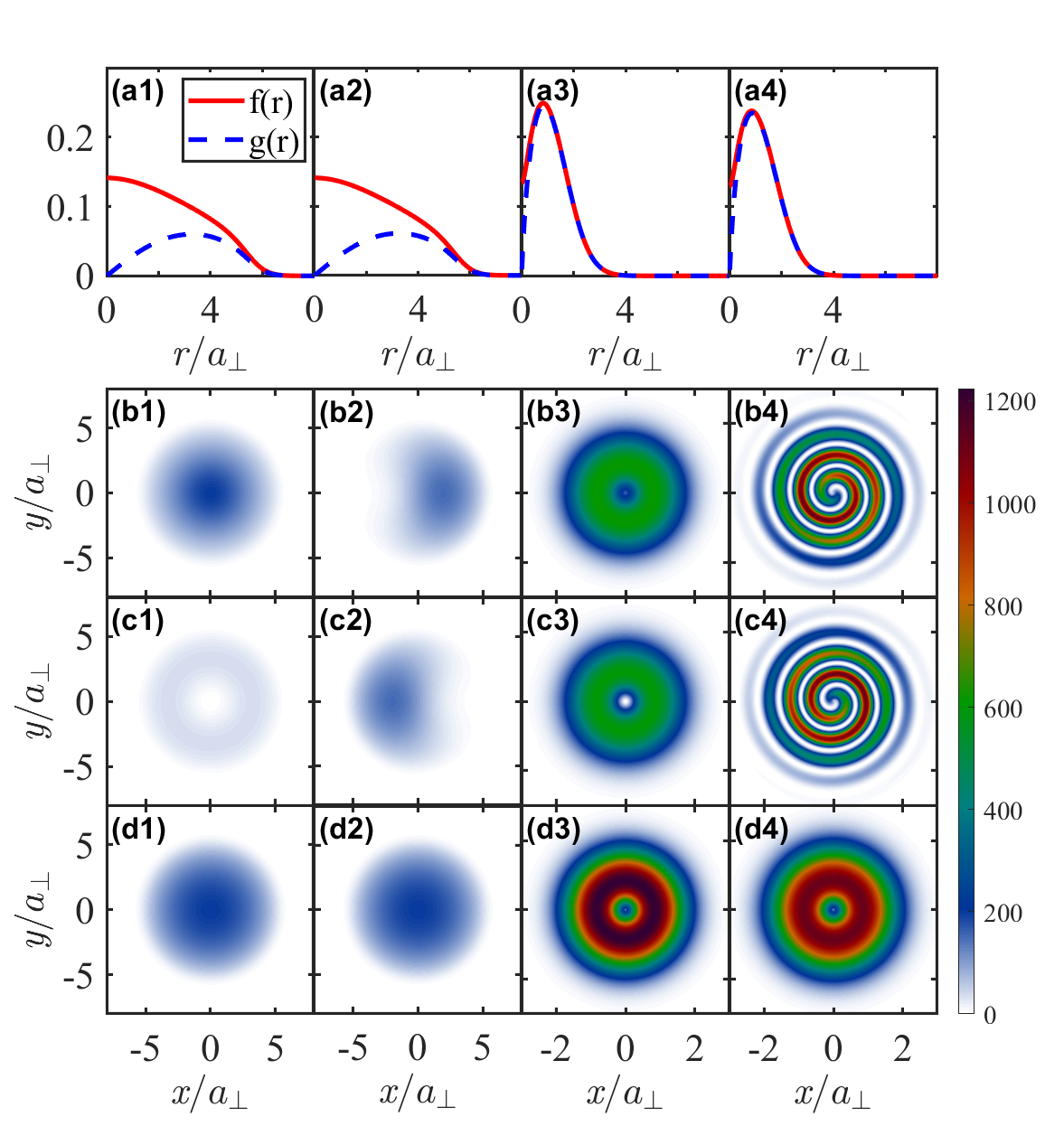}
\caption{(Color online) The spin and total particle densities, and the functions of $f(r)$ and $g(r)$ of four phase. (b1), (c1) and (d1) are spin up, spin down and total densities for HV phase, (a1) is the corresponding $f(r)$ and $g(r)$, where $k_0 a_{\perp}=0.1$, $Ng/\omega a_\bot^2=400,Ng^\prime/\omega a_\bot^2=399$; (a2), (b2), (c2) and (d2) are for DP phase, where $k_0 a_{\perp}=0.1$, $Ng/\omega a_\bot^2=400, Ng^\prime/\omega a_\bot^2=401$; (a3), (b3), (c3) and (d3) are for RWHV phase,  where $k_0 a_{\perp}=5$, $Ng/\omega a_\bot^2=7, Ng^\prime/\omega a_\bot^2=5$; (a4), (b4),(c4) and (d4) are for SS phase, where $k_0 a_{\perp}=5$, $Ng/\omega a_\bot^2=7, Ng^\prime/\omega a_\bot^2=9$.  }
\label{fig:density}
\end{figure}
Four types of states, HV, RWHV, DP and SS, can be classified according to the values of $k_c$ and $\Delta E$. The classification can be summarized in Tab. \ref{tab}.
\begin{table}[h]
\begin{center}
\begin{tabular}{|c|c|c|}\hline &
$k_c=0$ & $k_c\neq 0$ \\
\hline
$\Delta E>0$ & ~~~~HV~~~~ & ~~~~RWHV~~~~ \\
\hline $\Delta E<0$ & ~~~~DP~~~~ & ~~~~SS~~~~
\\ \hline
\end{tabular}
\end{center}
\caption{Four phases can be classified by the criterion of $k_c=0$ or $k_c\neq 0$, $\Delta E<0$ or $\Delta E>0$.}\label{tab}
\end{table}
%\diagbox{$\Delta E$}{phase}{$k_c$}&
In the state of HV the radial wave factor $e^{ik_cr}$ is absent since $k_c=0$. The spin densities are demonstrated in Fig. \ref{fig:density} (b1) and (c1). The spin down density has a vortex structure while the spin up density doesn't. This is why it's called HV and this state has been discussed in Ref.\cite{Wu2011,Ramachandhran2012}. Our calculation shows that the spin down component is severely suppressed and vortex structure is barely observable when $Ng^\prime/\omega a_\bot^2$ and $Ng/\omega a_\bot^2$ are small. Hence, in Fig. \ref{fig:density} (b1) and (c1) we demonstrate the HV structure with large $Ng^\prime/\omega a_\bot^2$ and $Ng/\omega a_\bot^2$ and they can be easily tuned by the total number density. The DP is a superposition state of HV and its time-reversal state. The densities of spin up and down component of this phase both show a peak structure, and the locations of the two peaks are mismatched as shown in Fig. \ref{fig:density} (b2) and (c2). This is the reason it's named as DP phase. The total particle density has single peak as shown in Fig. \ref{fig:density} (d2).  Of particular interest are the states of RWHV and SS. The RWHV also has a half vortex structure as shown in Fig. \ref{fig:density} (b3) and (c3). Moreover, in the RWHV state $k_c$ is nonzero, and thus it is a half vortex combined with a radial wave. However, this radial wave factor doesn't manifest itself in the density profile. The SS state is a superposition of RWHV and its time-reversal state. The superposition of factors $e^{ik_c r}$ and $e^{-ik_c r}$ leads to a periodic modulation of the spin density along the radial direction. Furthermore, this modulation also depend on the angle $\theta$. Eventually, the spin densities of SS phase demonstrate an intriguing spiral pattern as shown in Fig. \ref{fig:density} (b4) and (c4).

\begin{figure}[t]
\includegraphics[width=0.5\textwidth]{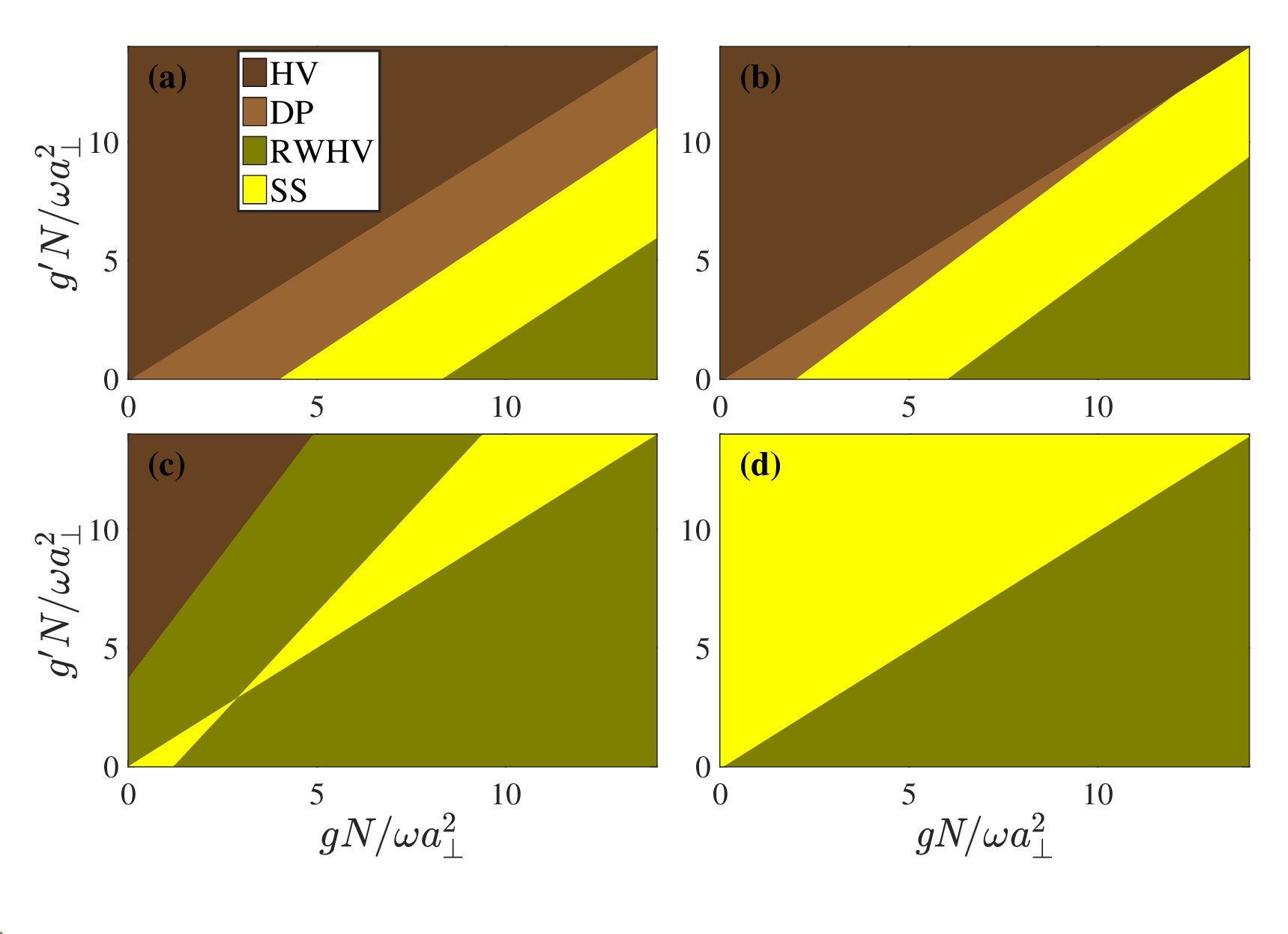}
\caption{(Color online) The phase diagrams for different SO coupling strength $k_0$. $k_0 a_{\perp}=0.1,0.5,1$ and $5$ for (a),(b),(c) and (d), respectively.}
\label{fig:pd}
\end{figure}
In Fig.\ref{fig:pd} (a),(b),(c) and (d) we demonstrate the phase diagrams for $k_0 a_{\perp}=0.1,0.5,1$ and $5$, respectively. From the phase diagrams one observes that the phases of HV and DP only exist for small $k_0$, the regions of HV and DP phases shrink and finally disappear as $k_0$ increases. For large SO coupling the phases of SS and RWHV are dominant. For instance, in Fig.\ref{fig:pd} (d), where $k_0 a_{\perp}=5$, only phases of SS and RWHV exist.  The SS phase appear in region $g^\prime>g$ while the RWHV phase takes the region of $g^\prime<g$ . As shown in Fig. \ref{fig:density} (d1) and (d2) the spin densities of SS phase have a spiral pattern. The high density region of one component overlaps with the low density region of the other component, and thus the interspecies interaction energy is lowered. Then the SS phase is more favored for $g^\prime>g$ region. Based on the diagrams, we may conclude that the phases with a radial wave factor $e^{ik_cr}$ are more favored for strong SO couplings.

\section{THE TOPOLOGICAL CHARGES AND SPIN TEXTURES}
The topological classification of the four phases can be quantified by the skyrmion topological charge,
\bea
Q=\frac{1}{4\pi}\int d^2 {\bf r} ~n_{\rm skyrmion}({\bf r}),
\eea
where $n_{\rm skyrmion}({\bf r})={\bf n}\cdot(\partial_x {\bf n}\times \partial_y {\bf n})$ is the skyrmion density, and ${\bf n({\bf r})}$ is a unit vector describing the direction of the spin, which can be calculated as
\bea
{\bf n({\bf r})}=\frac{\phi^\dagger_s\vec{\sigma}\phi_s}{\phi_s^\dagger \phi_s}.
\eea Skyrmion (anti-skyrmion) has a charge of $Q=1(-1)$, while the meron (anti-meron) is characterized by the charge of $Q=\frac{1}{2}(-\frac{1}{2})$.

\begin{figure}[t]
\includegraphics[width=0.47\textwidth]{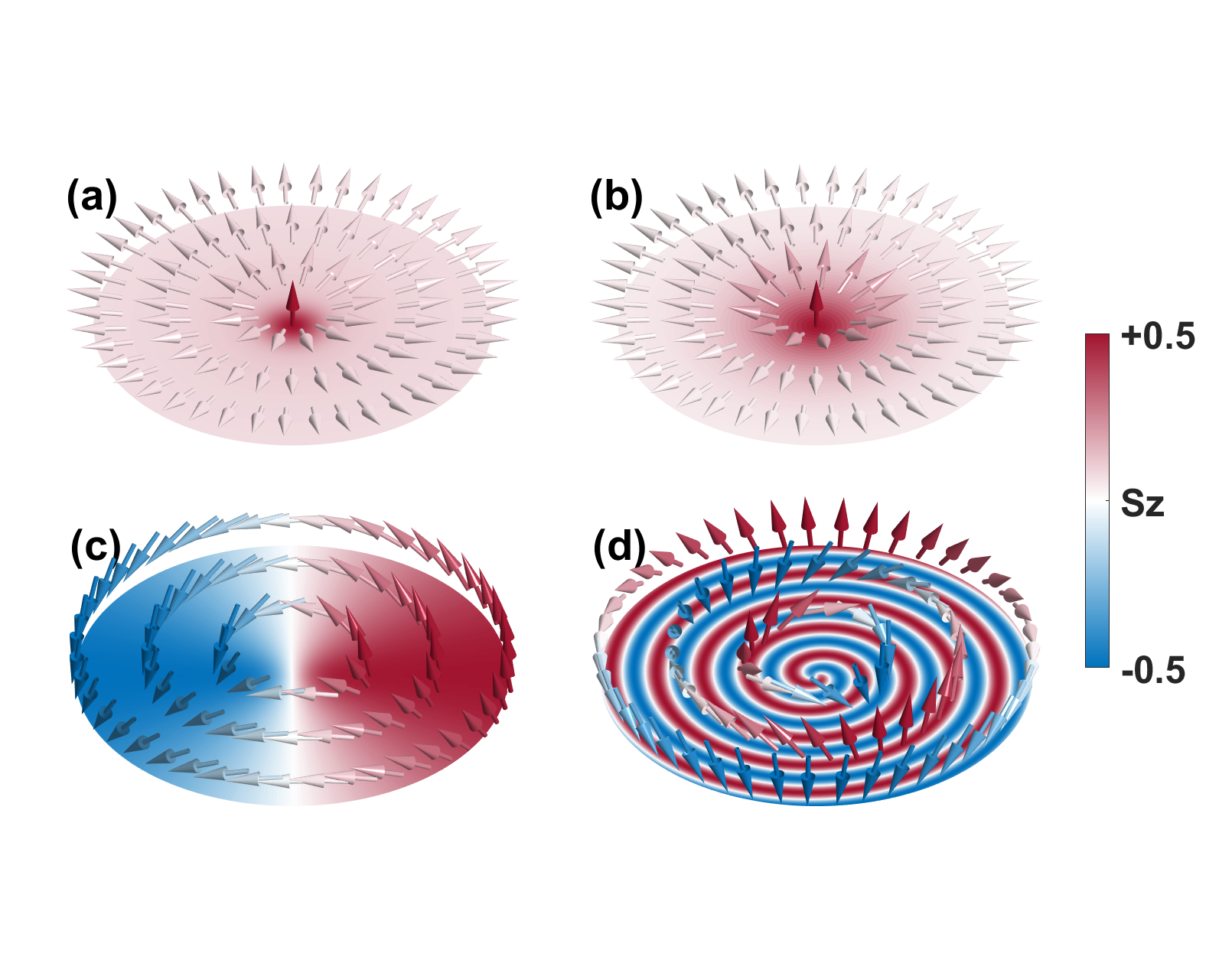}
\caption{(Color online) The spin distributions $\bf{S}({\bf r})$ for (a) HV phase, where $k_0 a_{\perp}=0.1$, $Ng/\omega a_\bot^2=400,Ng^\prime/\omega a_\bot^2=399$; (b) RWHV phase,  where $k_0 a_{\perp}=5$, $Ng/\omega a_\bot^2=7, Ng^\prime/\omega a_\bot^2=5$; (c) DP phase, where $k_0 a_{\perp}=0.1$, $Ng/\omega a_\bot^2=400, Ng^\prime/\omega a_\bot^2=401$, and (d) SS phase, where $k_0 a_{\perp}=5$, $Ng/\omega a_\bot^2=7, Ng^\prime/\omega a_\bot^2=9$. $\gamma=0$ for all the graphs.}
\label{fig:st}
\end{figure}

Straight-forward calculation yields
\begin{align}
&n_{\rm skyrmion}({\bf r})=\cr &\frac{f(r)g(r)}{\pi r(f^2(r)+g^2(r))^2}\big(f(r)\partial_rg(r)-g(r)\partial_r f(r)\big),\label{eq:SD1}
\end{align} for $|\alpha|=1$ and $\beta=0$, which are the cases of HV and RWHV. We evaluated the topological charge $Q$ using Eq. (\ref{eq:SD1}) at representative parameter points in Fig. \ref{fig:density} for the HV and RWHV phases. The numerical integration yields $Q\approx0.39$ for the HV phase and $Q=0.499$ for the RWHV phase. Both values are close to meron classification ($Q=0.5$). The deviation from $Q=0.5$ for HV is a numerical artifact. To ensure numerical accuracy, we define an effective boundary of the system at the radius where either $f(r)$ or $g(r)$ drops to $10^{-5}$, which is one order of magnitude larger than the convergence tolerance of $10^{-6}$ in our imaginary-time evolution method. All our calculations are performed within this effective boundary.  At this effective boundary the spins of HV are not fully in-plane (as seen in Fig. \ref{fig:st}), so the truncated integration does not cover the complete meron. In the RWHV phase the wave function is more tightly localized at the trap center compared to the HV case, as clearly seen in Fig. \ref{fig:density}. Then the numerical integral of the skyrmion density naturally covers the complete meron configuration. Consequently, the topological charge obtained for RWHV is essentially exact. The existence of an isolated meron is due to two features of this system. First, with the SO coupling effect the spins tend to swirl along the boundary, which is exactly the spin configuration of the meron. Second, this system is confined in a harmonic trap. An isolated meron can be hosted in a finite size system. At the boundary of the trap, the functions $f(r)$ and $g(r)$ approach zero.

\begin{figure}[t]
\includegraphics[width=0.47\textwidth]{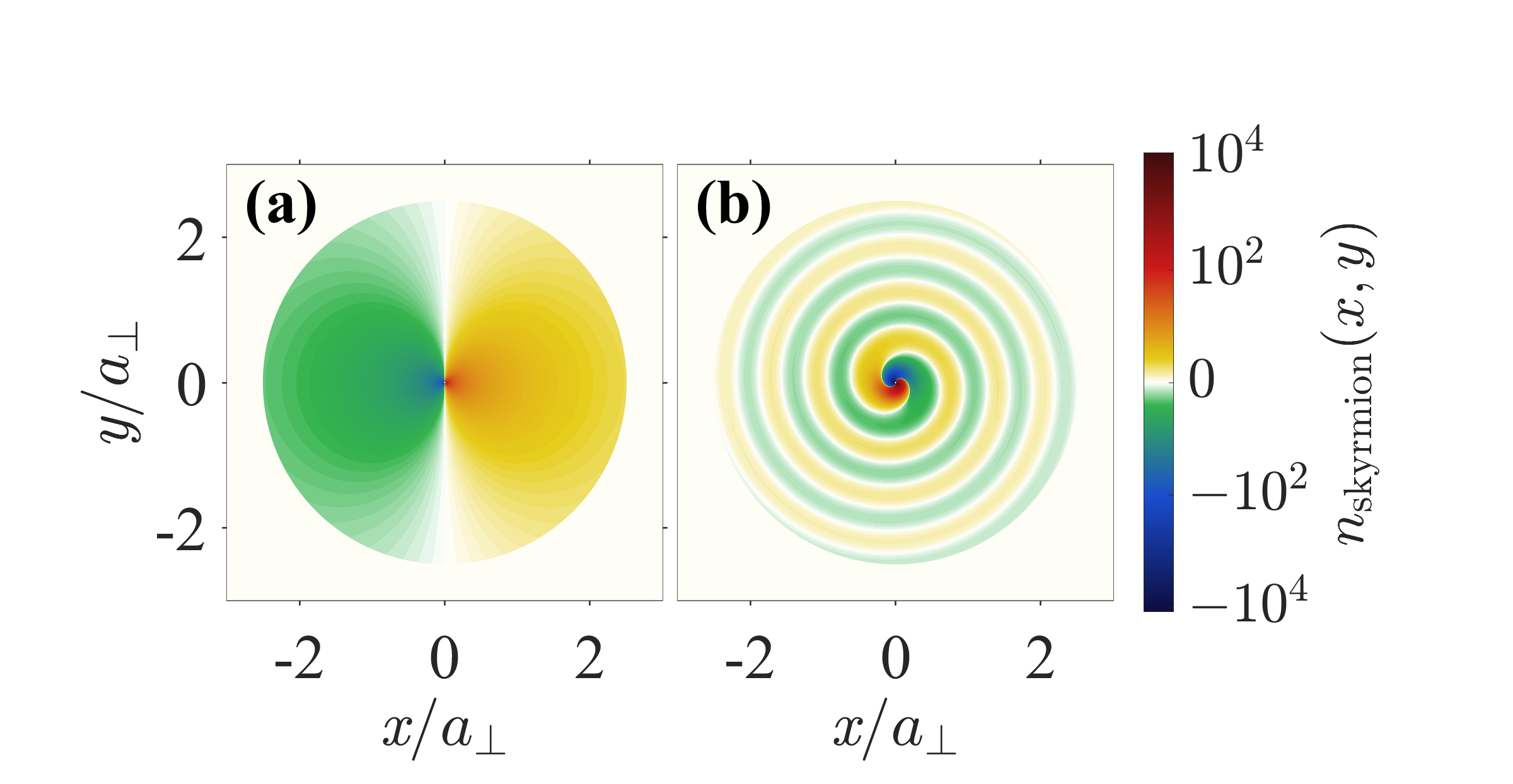}
\caption{(Color online) The skyrmion density $n_{\rm skyrmion}({\bf r})$ for (a) DP phase, where $k_0 a_{\perp}=0.1$, $Ng/\omega a_\bot^2=400, Ng^\prime/\omega a_\bot^2=401$, (b) SS phase, where $k_0 a_{\perp}=5$, $Ng/\omega a_\bot^2=7, Ng^\prime/\omega a_\bot^2=9$. $\gamma=0$ for (a) and (b).}
\label{fig:SD}
\end{figure}

The skyrmion density for $|\alpha|=|\beta|=1/\sqrt2$ is calculated as
\begin{align}
&n_{\rm skyrmion}({\bf r})=\cr &\frac{k f(r)g(r)}{\pi r(f^2(r)+g^2(r))}\sin(2kr+\theta+\gamma)\cr &+\frac{f(r)\partial_rg(r)-g(r)\partial_r f(r)}{\pi r(f^2(r)+g^2(r))^2}\cos(2kr+\theta+\gamma),
\end{align}
where $\gamma$ is the relative phase between $\alpha$ and $\beta$. This is the expression of skyrmion density for the DP and SS phases. It's straight-forward to see that the integration of the skyrmion density with respect to the angle $\theta$ yields zero topological charges for these two phases. Moreover, in Fig. \ref{fig:SD} we plot the skyrmion densities for DP and SS phases, which manifestly demonstrates the spatial distribution of positive and negative topological charge densities. It is clearly seen that the opposite charge densities cancel out each other, resulting in a zero total topological charge upon integration over the entire two-dimensional plane. Basically, these two states are superpositions of meron and antimeron as demonstrated in Eq. (\ref{eq:phis}), that is, they are the meronium states. It is crucial to distinguish between a generic meron-antimeron pair and a ``meronium" state. A meron-antimeron pair is a spatially separated configuration where the meron and antimeron retain distinct, non-overlapping centers. In contrast, we adopt the suffix ``-ium" for the overlapped configuration, where the meron and antimeron share an identical center of mass. In our system, the DP and SS phases belong to the latter class, as their wave functions manifest a superposition of two spin textures centered precisely at the trap center.

To gain more insight on topological properties of the four phases, it's useful to illustrate the spin distribution
\bea\bf{S}({\bf r})=\frac{1}{2}{\bf n({\bf r})}.\eea They
are plotted in Fig. \ref{fig:st}. The graph (a) and (b) are for states of HV and RWHV, respectively. They are typical meron configurations. The graph Fig. \ref{fig:st} (c) and (d) show the spin configurations of DP and SS phases. They are both meroniums. The spin configuration of the SS phase is more complicated due to the superposition of the factors of $e^{ik_c r}$ and $e^{-ik_c r}$. Usually, such a zero topological charge state is unstable because of the topologically triviality. From Fig. \ref{fig:pd} one observe that the DP state only appear in regions of small $k_0$ and $g\approx g^\prime$, while the region of SS phase in the phase diagram becomes larger as $k_0$ increase. Hence, one can conclude that the strong SO coupling effect tend to stabilize the meronium state of SS type, while destabilize the DP type.

\section{Conclusions}
The topological states in a two-dimensional Rashba SO coupled Bose gas are investigated. Systems with SO couplings usually tend to condense in a state with nonzero momentum. Analogously, we find that topological state in our system can have a radial wave factor $e^{ikr}$ in some parameter space. Basically, we find four distinguished phases. One is the HV phase, which has been found and discussed in similar systems. The second one is the RWHV, which is a HV combined with a radial wave factor $e^{ikr}$. The other two phases, DP and SS, are superpositions of HV and RWHV, respectively. Due to the superposition of $e^{ik_c r}$ and $e^{-ik_c r}$, the SS phase exhibits novel and intriguing structure, the spin density of which has a spiral pattern. Furthermore, the spin configurations show that HV and RWHV are merons. We also find that the spin configuration of DP and SS states are new type, called meronium. Phase diagram shows that the DP is less stable, while SS phase is more stable for strong SO coupling effect. In material science,the combinations of topological structures with positive and negative charges may have broader applications due to the absence of skyrmion Hall effect. Hence, this work has not only uncovered a new topological state, but also provides insights into the search for combined topological states in material science. In future work, it will be interesting to systematically investigate the instability of the meronium state under various circumstances, for instance, the instability to higher order angular momentum states, against the anisotropy of the SO coupling strength or anisotropy of the trap potential.

\section{Acknowledgement}
The authors thank Wu-Ming Liu for useful discussions. The work is supported by the National Science Foundation of China (Grant No. NSFC-11874002).

\appendix
\begin{widetext}
\section{Derivations of the energy functional and Euler-Lagrange equations}
In the polar coordinates the Hamiltonian $\hat H_0$ can be expressed as
\bea
\hat H_{0}=\int d^2 {\bf r}\hat\Psi^\dagger({\bf r})\mathcal{\hat H}_0\hat\Psi(\bf r),
\eea where
\bea
\mathcal{\hat H}_0= \begin{bmatrix}
-\dfrac{1}{2m} \left( \dfrac{\partial^2}{\partial r^2} + \dfrac{1}{r}\dfrac{\partial}{\partial r} + \dfrac{1}{r^2}\dfrac{\partial^2}{\partial \theta^2} \right) & \dfrac{ik_0}{m} e^{-i\theta} \left( \dfrac{\partial}{\partial r} - \dfrac{i}{r}\dfrac{\partial}{\partial \theta} \right) \\[10pt]
\dfrac{ik_0}{m} e^{i\theta} \left( \dfrac{\partial}{\partial r} + \dfrac{i}{r}\dfrac{\partial}{\partial \theta} \right) & -\dfrac{1}{2m} \left( \dfrac{\partial^2}{\partial r^2} + \dfrac{1}{r}\dfrac{\partial}{\partial r} + \dfrac{1}{r^2}\dfrac{\partial^2}{\partial \theta^2} \right)
\end{bmatrix}
\eea
Using the mean-field condensate wave function $\Psi=\sqrt N \phi_n$ the energy functional can be directly calculated as
\begin{align}
&E_0(\phi_n)= N \int d^2{\bf r} \Big\{ -\frac{1}{2m} \Big[ f(r) \frac{\partial^2 f(r)}{\partial r^2} + \frac{f(r)}{r} \frac{\partial f(r)}{\partial r} - \left( k^2 + \frac{n^2}{r^2} \right) f^2(r) + g(r) \frac{\partial^2 g(r)}{\partial r^2} + \frac{g(r)}{r} \frac{\partial g(r)}{\partial r}\nonumber\\&- \left( k^2 + \frac{(n+1)^2}{r^2} \right) g^2(r) \Big]- \frac{2kk_0}{m} f(r)g(r)\nonumber
\\&-\frac{ik}{m} \left( f(r) \frac{\partial f(r)}{\partial r}+ \frac{1}{2r} f^2(r)  + g(r) \frac{\partial g(r)}{\partial r} + \frac{1}{2r} g^2(r) \right)
 + \frac{ik_0}{m} \left( f(r) \frac{\partial g(r)}{\partial r}  + g(r) \frac{\partial f(r)}{\partial r} +\frac{1}{r} f(r)g(r) \right) \Big\}.\label{eq:appE0}
\end{align}
The energy of a Hermitian system should be real. The imaginary parts in above energy functional can be written as a total derivative. For instance,
\begin{align}
\int d^2{\bf r} \left(f(r)\frac{\partial f(r)}{\partial r}+ \frac{1}{2r}f^2(r)\right) =2\pi \int_{0}^{\infty}dr\left(rf(r)\frac{\partial f(r)}{\partial r}+ \frac{1}{2}f^2(r)\right) =\pi\int_{0}^{\infty}dr\partial_r\left(rf^2(r)\right)=\pi rf^2(r)\mid_0^\infty.
\end{align}
For a system in trap, the wave functions approach zero as $r\longrightarrow\infty$ and hence above integration vanishes.
By the same token,$\int d^2{\bf r} \left(g(r)\frac{\partial g(r)}{\partial r}+ \frac{1}{2r}g^2(r)\right)=0$, and $\int d^2{\bf r} \left( f(r) \frac{\partial g(r)}{\partial r}  + g(r) \frac{\partial f(r)}{\partial r} +\frac{1}{r} f(r)g(r) \right)=0$.

Integrating by parts, the term $\int d^2{\bf r} \left( f(r) \frac{\partial^2 f(r)}{\partial r^2}\right)$ in Eq.(\ref{eq:appE0}) can be calculated as
\begin{align}
\int d^2{\bf r} \left( f(r) \frac{\partial^2 f(r)}{\partial r^2}\right)&=2\pi\int_{0}^{\infty} dr rf(r) \frac{\partial}{\partial r} \frac{\partial f(r)}{\partial r} = 2\pi\frac{\partial f(r)}{\partial r} rf(r)\bigg|_{0}^{\infty} -2\pi \int_{0}^{\infty} dr \left[ \frac{\partial f(r)}{\partial r} \left( f(r) + r \frac{\partial f(r)}{\partial r} \right) \right]\nonumber\\&= - 2\pi\int_{0}^{\infty} dr \left( f(r)\frac{\partial f(r)}{\partial r}   + r \left(\frac{\partial f(r)}{\partial r} \right)^2 \right).
\end{align} Analogously,
\bea
\int d^2{\bf r} \left( g(r) \frac{\partial^2 g(r)}{\partial r^2}\right) = - 2\pi\int_{0}^{\infty} dr \left( f\frac{\partial g(r)}{\partial r}   + r \left(\frac{\partial g(r)}{\partial r} \right)^2 \right). \eea Taking these two integrations back in Eq.(\ref{eq:appE0}) the energy functional $E_0(\phi_n)$ can be expressed as
\begin{align}
E_0(\phi_n)&= N\int d^2 {\bf r}\Big\{\frac{1}{2m}\big[(\partial_r f(r))^2+(\partial_r g(r))^2+\left(k^2+\frac{n^2}{r^2}\right)f^2(r)+\left(k^2+\frac{(n+1)^2}{r^2}\right)g^2(r)-4kk_0f(r)g(r)\big]\Big\}.
\end{align}

In the mean-field level, it's straight-forward to calculate the interaction part as
\begin{align}
E_{\text{int}} &= N^2 \int d^2 {\bf r} \left[ g(f^4(r) + g^4(r)) +2g^\prime f^2(r) g^2(r) \right]\nonumber\\&=N^2 \int d^2 {\bf r} \left[ \frac{g+g^\prime}{2} (f^2(r) + g^2(r))^2 +\frac{g-g^\prime}{2} (f^2(r) - g^2(r)) ^2 \right].
\end{align}
Including the trap potential part, the total energy functional can be written as
\begin{align}
E(\phi_n)&=E_0(\phi_n)+E_{int}+ N\int d^2 {\bf r}\frac{1}{2}m\omega^2r^2(f^2(r)+g^2(r))\nonumber\\ &+N\int d^2 {\bf r}\Big\{\frac{1}{2m}\big[(\partial_r f(r))^2+(\partial_r g(r))^2+\left(k^2+\frac{n^2}{r^2}\right)f^2(r)+\left(k^2+\frac{(n+1)^2}{r^2}\right)g^2(r)-4kk_0f(r)g(r)\big]\nonumber\\&
+\frac{1}{2}m\omega^2r^2(f^2(r)+g^2(r))+(g+g^\prime)\frac{N}{2}(f^2(r)+g^2(r))^2+(g-g^\prime)\frac{N}{2}(f^2(r)-g^2(r))^2\Big\}.\label{eq:E}
\end{align}

To find the energy minimum with the constraint $\int d^2 {\bf r}(f^2+g^2)=1$ we employ the Lagrange multiplier method, that is, we introduce a chemical potential term to the energy $E^\prime(\phi_n)= E(\phi_n)-\mu \int d^2 {\bf r}(f^2+g^2)$. Then the Euler-Lagrange Equations can be derived as
\begin{align}
&\frac{\partial}{\partial r}\left(\frac{\partial I }{\partial(\partial_r f(r))}\right)=\frac{\partial I}{\partial f(r)}\nonumber\\
&\frac{\partial}{\partial r}\left(\frac{\partial I }{\partial(\partial_r g(r))}\right)=\frac{\partial I}{\partial g(r)},
\end{align}
where $I$ is the integrand of the energy $E^\prime(\phi_n)$,
\begin{align}
I&= 2\pi N r\Big[\frac{1}{2m}\big[(\partial_r f(r))^2+(\partial_r g(r))^2+\left(k^2+\frac{n^2}{r^2}\right)f^2(r)+\left(k^2+\frac{(n+1)^2}{r^2}\right)g^2(r)-4kk_0f(r)g(r)\big]\nonumber\\&
+\frac{1}{2}m\omega^2r^2(f^2(r)+g^2(r))+(g+g^\prime)\frac{N}{2}(f^2(r)+g^2(r))^2+(g-g^\prime)\frac{N}{2}(f^2(r)-g^2(r))^2-\mu(f^2(r)+g^2(r))\Big].
\end{align}
Then, we obtain the differential equations of $f(r)$ and $g(r)$ as
\begin{align}
&\frac{1}{2m}\left(-\frac{\partial^2 f(r)}{\partial r^2}-\frac{1}{r}\frac{\partial f(r)}{\partial r}+\left(k^2+\frac{n^2}{r^2}\right)f(r)\right)-\frac{kk_0}{m}g(r)+\frac{1}{2}m\omega^2r^2f(r)+N(g+g^\prime)\big(f^2(r)+g^2(r)\big)f(r)\cr&+N(g-g^\prime)\big(f^2(r)-g^2(r)\big)f(r)-\mu f(r)=0, \cr&\frac{1}{2m}\left(-\frac{\partial^2 g(r)}{\partial r^2}-\frac{1}{r}\frac{\partial g(r)}{\partial r}+\left(k^2+\frac{(n+1)^2}{r^2}\right)g(r)\right)-\frac{kk_0}{m}f(r)+\frac{1}{2}m\omega^2r^2g(r)+N(g+g^\prime)\big(f^2(r)+g^2(r)\big)g(r)\cr&+N(g-g^\prime)\big(f^2(r)-g^2(r)\big)g(r)-\mu g(r)=0.
\end{align}
\end{widetext}

\end{document}